\documentclass[12pt,a4paper]{article}
\usepackage{amsmath}
\usepackage{amsxtra}
\usepackage{amstext}
\usepackage{amssymb}
\usepackage{latexsym}
\usepackage{dsfont}
\usepackage{graphics}

\newcommand{\nn}{\nonumber}

\newcommand{\p}{\vspace{6pt}\noindent}
\newcommand{\jump}{\vspace{2pt}}

\advance\textheight by \topskip
\makeatletter

\@addtoreset{equation}{section}
\def\section{\@startsection {section}{1}{\z@}{-8.5ex plus -1ex minus
 -.2ex}{3.3ex plus .2ex}{\large\bf}}%\centering}}
\def\subsection{\@startsection{subsection}{2}{\z@}{-3.25ex plus
 -1ex minus -.2ex}{1.5ex plus .2ex}{\bf}}
\def\subsubsection{\@startsection{subsubsection}{3}{\z@}{-3.25ex plus%
 -1ex minus -.2ex}{1.5ex plus .2ex}{\sl}}

\begin{document}
\begin{titlepage}
\vspace*{-2cm}
\begin{flushright}
%%YITP-05-29
\end{flushright}

\vspace{0.3cm}

\begin{center}
{\Large {\bf }} \vspace{1cm} {\Large {\bf Defects in affine
Toda field theories}}\\
\vspace{0.3cm}
C.\ Zambon\footnote{\noindent E-mail: {\tt cristina.zambon@ptm.u-cergy.fr}} \\
\vspace{0.3cm}
{\em Laboratoire de Physique Th\'{e}orique et Mod\'{e}lisation \\
Universit\'{e} de Cergy-Pontoise (CNRS UMR 8089), Saint-Martin 2\vspace{.1cm}\\
2 avenue Adolphe Chauvin,
95302 Cergy-Pontoise Cedex, France}\\

\vspace{1cm} {\bf{ABSTRACT}}
\end{center}

\p In this talk some classical and quantum aspects concerning a
special kind of integrable defect - called a jump-defect - will be
reviewed. In particular, recent results obtained in an attempt to
incorporate this defect in the affine Toda field theories, in
addition to the sine-Gordon model, will be presented.

\vfill
\end{titlepage}

\section{Jump-defects}
\label{freefield}

\p The jump-defect is a purely transmitting defect, which can be
incorporated in certain integrable field theories in such a way as
to allow the integrability of the system to be preserved. Its
existence has been proved originally for the sine-Gordon model, both
in the classical \cite{bczlandau} and in the quantum context
\cite{bczsg05}, and subsequently extended to other integrable
systems (see for instance \cite{Corrigan06}). From the very start,
the jump-defect has displayed interesting features, which appear to
be quite different from the ones enjoyed by a typical $\delta$-type
impurity.

\p Consider two free massive scalar fields $\phi(x,t)$, $x<0$ and
$\psi(x,t)$, $x>0$, with Lagrangian density given by
\begin{equation}\label{DefectLag}
{\cal L}=\theta(-x){\cal L}_\phi +\theta(x){\cal L}_\psi+
\delta(x){\cal D}(\phi,\psi).
\end{equation}
The terms ${\cal L}_\phi$ and ${\cal L}_\psi$ represent the bulk
Lagrangian densities ($-\infty < x< \infty$) for the fields $\phi$
and $\psi$ respectively, while ${\cal D}$ defines the condition for
the defect, which is located at $x=0$. The function ${\cal D}$ can
be chosen in many ways. For instance, for a $\delta$-type impurity
\begin{equation}\label{DefectTermDelta}
{\cal D}=-\frac{1}{2}\left[\sigma\, \phi\psi -(\phi_x+\psi_x)(\phi -
\psi )\right],
\end{equation}
which leads to the following set of equations of motion and defect
conditions
\begin{eqnarray}\label{DefectConditionFreeDelta}
&\partial^2\phi=-m^2 \quad x<0, \qquad &\phi=\psi \quad
x=0, \nn\\
&\partial^2\psi=-m^2 \quad x>0, \qquad &\psi_x-\phi_x=\sigma \phi
\quad x=0.
\end{eqnarray}
Because of the presence of a defect, which breaks the space
translation invariance, it can be verified that momentum is not
conserved. Moreover, the system described by the Lagrangian density
\eqref{DefectLag} with defect term \eqref{DefectTermDelta} allows
both transmission and reflection and possesses a bound state,
provided $\sigma<0$.

\p On the other hand, a defect term, which defines the condition for
a jump-defect reads
\begin{equation}\label{DefectTermJump}
{\cal D}=\frac{1}{2}\left(\phi \psi_t- \psi
\phi_t\right)+\frac{m\,\sigma}{4}(\phi+\psi)^2
+\frac{m}{4\,\sigma}(\phi-\psi)^2.
\end{equation}
As a consequence, the Lagrangian density \eqref{DefectLag} with
${\cal D}$ given by \eqref{DefectTermJump} leads to the following
set of equations
\begin{eqnarray}\label{DefectConditionFreeJump}
&\partial^2\phi=-m^2 \quad x<0, \qquad &\partial_x\phi -\partial_t
\psi = -\sigma\left(\frac{\phi + \psi }{2}\right)-\frac{1}{\sigma}
\left(\frac{\phi - \psi }{2}\right)\quad
x=0, \nn\\
&\partial^2\psi=-m^2 \quad x>0, \qquad &\partial_x \psi
-\partial_t\phi = \phantom{-}\sigma\left(\frac{\phi + \psi
}{2}\right)-\frac{1}{\sigma}\left(\frac{\phi - \psi }{2}\right)\quad
x=0.
\end{eqnarray}
This time, surprisingly, it was discovered that momentum is
conserved \cite{bczlandau}, provided a suitable contribution from
the defect is added. Besides, the system described by equations
\eqref{DefectConditionFreeJump} is purely transmitting and does not
have a bound state. Finally, it is worth pointing out that contrary
to the $\delta$-type impurity situation
\eqref{DefectConditionFreeDelta}, if there is a jump-defect, the two
fields $\phi$ and $\psi$ do not match at the defect location.

\p This very simple example is useful to elucidate some of the more
striking features of a jump-defect, which persist when it is
incorporated in some integrable field theory such as the sine-Gordon
model. In addition, contrary to a $\delta$-type impurity, the
jump-defect is able to preserve integrability.

\section{Jump-defects and affine Toda field theories}

\p The aim of this talk is to summarize some recent results obtained
in the context of the affine Toda field theories (ATFTs)
\cite{Arinshtein79} (see also the review \cite{Corrigan94} and
references therein) with a jump-defect, namely the progress achieved
in stretching the investigation beyond the sine-Gordon model. For
this purpose, only the ATFTs associated with the root data of the
Lie algebra $a_r$ will be taken into account. Classically,
integrable jump-defects incorporated into these models have been
described extensively in \cite{bcztoda}, where their integrability
was established via a Lax pair argument. However, only recently, it
has been possible to add a quantum description of the $a_2$ affine
Toda model with a jump-defect \cite{Corrigan07}.

\subsection{Classical setting}
\label{Classical Toda}

\p The bulk Lagrangian density for a complex ATFT with root data of
the Lie algebra $a_r$ reads
\begin{equation}\label{BulkDenLag}
{\cal L}_{\phi} =\frac{1}{2} (\partial_{\mu}\phi\cdot
\partial^{\mu}\phi)+\frac{m^{2}}
{\beta^{2}}\,\sum_{j=0}^{r}\,
(e^{i\beta\alpha_{j}\cdot\phi}-1),\qquad |\alpha_j|^2=2
\end{equation}
where $m$ and ${\beta}$ are constants, and $r$ is the rank of the
algebra. The vectors $\alpha_j$ with $j=1,\dots,r$ are simple roots
and $\alpha_0$ is the extended root, defined by
$$\alpha_0=-\sum_{j=1}^{r}\,\alpha_j.$$
The field $\phi=(\phi_1,\phi_2,\, \dots,\, \phi_r)$ takes values in
the $r$-dimensional Euclidean space spanned by the simple roots
$\{\alpha_j\}$. The ATFTs described by the Lagrangian density
\eqref{BulkDenLag} are massive integrable field theories. They
possess infinitely many conserved charges, a Lax pair
representation, and many other interesting properties, both in the
classical and quantum domains. The simplest choice $r=1$ coincides
with the sine-Gordon model. Apart this model, all other ATFTs
described by the Lagrangian density \eqref{BulkDenLag} are not
unitary theories.

\p What is particular interesting in the context of the present
investigation is the fact that these models possess soliton
solutions \cite{Hollowood92}, for which the explicit expression
reads
\begin{equation}\label{todasoliton}
\phi^{a}=\frac{m^2i}{\beta}\sum^{r}_{j=0}\alpha_{j}\ln \left(1+
E_a\,\omega^{aj}\right)\quad a=1,\dots,r, \qquad E_a=e^{a_a x-b_a t
+\xi_a},\qquad \omega=e^{2\pi i/h},
\end{equation}
where $(a_a, b_a)=m_{a}\, (\cosh{\theta},\sinh{\theta})$, $h=(r+1)$
is the Coxeter number of the algebra, $\xi_a$ is a complex
parameter, and $\theta$ is the soliton rapidity. These soliton
solutions are complex, with the exception for the sine-Gordon
soliton. Nevertheless, they possess real energy and momentum, and
their masses are given by \cite{Hollowood92}
\begin{equation}\label{singlesolitonmasses}
M_a=\frac{4\,h\,m}{\beta^2}\sin\left(\frac{\pi a }h\right).
\end{equation}
Each solution \eqref{todasoliton} is characterized by a topological
charge, which is defined to be
\begin{equation}\label{topologicalcharges}
Q^{a}=\frac{\beta}{2\pi}\int^{\infty}_{-\infty}dx\,\partial_x\phi^{a}=
\frac{\beta}{2\pi}\phi^{a}(\infty,t),
\end{equation}
which lies in the weight lattice $\Lambda_W(a_r)$ of the Lie algebra
$a_r$. In particular, it can be noticed that for each $a=1,\dots, r$
there are several solitons whose topological charges lie in the set
of weights of the fundamental $a^{th}$ representation of $a_r$
\cite{McGhee94}. Looking at the expression \eqref{todasoliton}, it
can be noticed that the value of the topological charge depends on
the imaginary part of parameter $\xi_a$. Shifting $\xi_a$ by $2\pi
ia/h$ changes the topological charge, since that amount sets the
boundaries between different topological charge sectors.

\p The system with a single jump-defect located at $x=0$, which
links two $a_r$ fields $\phi(x,t)$, $x<0$ and $\psi(x,t)$, $x>0$, is
described by the Lagrangian density \eqref{DefectLag} with defect
term given by
\begin{equation}\label{DefectTermToda}
{\cal D}=\left(\frac{1}{2}\phi\cdot E\partial_t\phi +\phi\cdot
D\partial_t\psi+\frac{1}{2}\psi \cdot E\partial_t\psi-{\cal B}(\phi,
\psi)\right).
\end{equation}
The requirement that integrability must be preserved, forces the
matrices $E$ to be antisymmetric with $D=1-E$ and fixes the form of
the defect potential ${\cal B}$ to be
\begin{equation}\label{DefectPotentialToda}
{\cal B}=-\frac{m}{\beta^2}\,\sum_{j=0}^r\left(\sigma\,
e^{i\beta\alpha_j\cdot(D^T\phi+D\psi)/2}+ \frac{1}{\sigma}\,
e^{i\beta\alpha_j \cdot D(\phi -\psi)/2}\right),
\end{equation}
where $\sigma$ represent the defect parameter. Moreover, the matrix
$D$ satisfies the following constraints
\begin{equation}\label{constraintsonD}
\alpha_k\cdot D\alpha_j=\left\{%
\begin{array}{ll}
    \phantom{-}2 & \hbox{$k=j$,} \\
    -2 & \hbox{$k=\pi(j)$, }\\
    \phantom{-}0 & \hbox{otherwise,} \\
\end{array}%
\right.\qquad (D+D^T)=2,
\end{equation}
where $\pi(j)$ indicates a permutation of the simple roots. Note
that for $r=1$, and after setting $\alpha_1=1/\beta=\sqrt{2}$, the
linearized version of \eqref{DefectTermToda} and
\eqref{DefectPotentialToda} reduce to \eqref{DefectTermJump}. It
should be mentioned that the jump-defect setting presented in this
section is not unique, as was explained in \cite{bcztoda}. However,
the alternative case will not be considered here.

\p Choosing a particular cyclic permutation, namely
$$
\alpha_{\pi(j)}=\alpha_{j-1}\quad\ j=1,\dots, r,\qquad
\alpha_{\pi(0)}=\alpha_r,
$$
it is possible to write explicitly the matrix $D$ as follows
\begin{equation}
D=2\sum_{j=1}^r w_j\left( w_j-w_{j+1}\right)^T, \qquad w_0\equiv
w_{r+1}=0,
\end{equation}
where $w_j$ with $j=1,\dots ,r$ are the fundamental highest weights
of the Lie algebra $a_r$ ($\alpha_i\cdot w_j=\delta_{ij}$). The
Lagrangian density \eqref{DefectLag} with defect term
\eqref{DefectTermToda} leads to the following equations of motion
\begin{equation}\label{equationsofmotion}
\partial^{2}\phi=\frac{m^2i}{\beta}\,
\sum_{j=0}^{r}\,\alpha_j\,e^{i\beta\alpha_{j}\cdot\phi}\quad x<0,
\qquad
\partial^{2}\psi=\frac{m^2i}{\beta}\,
\sum_{j=0}^{r}\,\alpha_j\,e^{i\beta\alpha_{j}\cdot\psi}\quad x>0,
\end{equation}
and defect conditions
\begin{equation}\label{defectconditions}
\partial_{x}\phi-E \partial_t \phi-D \partial_t \psi=0-\partial_\phi {\cal B}
 \quad x=0, \qquad
\partial_{x}\psi-D^{T}\partial_t \phi +E \partial_t \psi
=0\partial_\psi {\cal B} \quad x=0.
\end{equation}
As already pointed out in the case of the free massive field in
section (\ref{freefield}), a generalized momentum is conserved.
Again, the system allows only transmission, and it is instructive to
look at what happens when a soliton solution $\phi^{a}$ ($x<0$)
\eqref{todasoliton} travels across the jump-defect from the left to
the right ($\theta >0$). As expected, the emerging soliton
$\psi^{a}$ ($x>0$) will experience a delay since its form will be
\begin{equation}
\psi^{a}=\frac{m^2 i}{\beta}\sum^{r}_{j=0}\alpha_{j}\ln (1+
z_a\,E_a\,\omega^{aj}),
\end{equation}
where the explicit expression for the delay $z_a$ is provided by the
defect conditions \eqref{defectconditions}, namely
\begin{equation}\label{delay}
z_a=\left(\frac{i\,e^{-(\theta-\eta)}+i e^{-i\gamma_a}}
{e^{-(\theta-\eta)}+i e^{i\gamma_a}}\right), \qquad
\gamma_a=\frac{\pi \,a}{h},\qquad \sigma=e^{-\eta}.
\end{equation}
This expression is in general complex and diverges when
\begin{equation}\label{poleClasSet}
\theta=\eta+\frac{i\pi}{2}\left(1-\frac{2 a}{h}\right).
\end{equation}
However, for the self-conjugate soliton $a=h/2$ (provided $r$ is
odd), the delay becomes real and coincides with the delay for the
sine-Gordon model \cite{bczlandau}. When this happens the soliton
can be absorbed by the defect since the pole \eqref{poleClasSet}
appears for a real value of the rapidity, namely $\theta=\eta$.
Finally, in \cite{bcztoda} it was also pointed out that a soliton
might be turned into one and only one of the adjacent solitons by
the jump-defect, provided the argument of the delay \eqref{delay} is
sufficiently large. In fact, the argument is given by
\begin{equation}\label{argdelay}
\tan(\mbox{arg}\, z_a)=-\left(\frac{\sin
2\gamma_a}{e^{-2(\theta-\eta)} + \cos2\gamma_a}\right),
\end{equation}
and therefore the phase shift produced by the defect can vary
between zero (as $\theta\rightarrow -\infty$) and $-2\gamma_a$ (as
$\theta \rightarrow \infty$) allowing a change in the topological
charge of the incoming soliton, since, as pointed out before, the
topological charge sectors are separated exactly by $2\gamma_a$.

\subsection{Quantum domain}
\label{Quantum domain}

\p In this section, recent developments concerning the quantization
of the $a_r$ ATFTs will be presented. In particular, the example
elucidated in this talk concerns the $a_2$ affine Toda model, for
which a complete analysis has been carried out in \cite{Corrigan07}.
The purpose of that investigation was to find the transmission
matrices, describing the interaction amongst a jump-defect and the
soliton and antisoliton solutions of the model. Two different
approaches were used for this purpose and they will be sketched in
the next section. Both methods make use of the assumption that the
topological charge of the system containing two $a_2$ fields $\phi$
($x<0$), $\psi$ ($x>0$) and the jump-defect located in $x=0$ is
conserved. This fact relies on the classical investigation, briefly
presented in section \ref{Classical Toda}, which suggests that both
solitons and defects carry a topological charge that can be
exchanged due to their mutual interaction.

\p Both procedures allow to determine the transmission matrices up
to an overall function of the rapidity. The first method consists of
a functional integral approach, which makes use of the Lagrangian
density \eqref{DefectLag} with defect term \eqref{DefectTermToda},
together with a bootstrap procedure. The second method consists in
solving directly the triangular equations, which represent a set of
consistency conditions among the bulk scattering $S$-matrices and
the unknown transmission matrices. Primary ingredients needed for
both investigations are the $S$-matrices for the $a_2$ ATFT.
Together with the $S$-matrices for the other $a_r$ models (with the
exception of the sine-Gordon model), they have been conjectured by
Hollowood \cite{Hollowood93}, who made use of the trigonometric
solutions of the Yang-Baxter equation found originally by Jimbo
\cite{Jimbo89} (see also references therein).

\p The Lie algebra $a_2$ has two fundamental representations, and
the weights belonging to the first representation can be written in
terms of simple roots as follows
\begin{equation}\label{weightsa2}
l_1=\frac{1}{3}(2\alpha_1+\alpha_2),\qquad
l_2=-\frac{1}{3}(\alpha_1-\alpha_2),\qquad
l_3=-\frac{1}{3}(\alpha_1+2\alpha_2).
\end{equation}
As a consequence this representation contains three solitons, while
the corresponding antisolitons have weights which are the negative
of these and lie in the second representation. The knowledge of the
soliton-soliton $S$-matrix suffices since it can be used to derive
the other $S$-matrices, which describe the interactions
soliton-antisoliton and antisoliton-antisoliton, by means of a
bootstrap procedure. Having said that, the soliton-soliton
$S$-matrix for the $a_2$ model can be written in the following
explicit form
\begin{equation}\label{SMatrix}
S{^{mn}_{kl}}\,(\theta_{12})=R{^{mn}_{kl}}\,(x_{12})\,\rho(\theta_{12}),\qquad
\theta_{12}=(\theta_1-\theta_2),\qquad x_{12}=\frac{x_1}{x_2},
\end{equation}
where $k$, $l$ label the incoming particles and $m$, $n$ label the
outgoing particles in a two-body scattering process, with the
particle $k$, $n$ having rapidity $\theta_1$, and the particle $l$,
$m$ having rapidity $\theta_2$. The explicit form for the $R$-matrix
is
\begin{eqnarray}\label{RMatrix}
R(x_{12})=\left(
                \begin{array}{ccccccccc}
                  a(x_{12}) & 0 & 0 & 0 & 0 & 0 & 0 & 0 & 0 \\
                  0 & x_{12}^{1/3}\,c & 0 & b(x_{12}) & 0 & 0 & 0 & 0 & 0 \\
                  0 & 0 & x_{12}^{-1/3}\,c & 0 & 0 & 0 & b(x_{12}) & 0 & 0 \\
                  0 & b(x_{12}) & 0 & x_{12}^{-1/3}\,c & 0 & 0 & 0 & 0 & 0 \\
                  0 & 0 & 0 & 0 & a(x_{12}) & 0 & 0 & 0 & 0 \\
                  0 & 0 & 0 & 0 & 0 & x_{12}^{1/3}\,c & 0 & b(x_{12}) & 0 \\
                  0 & 0 & b(x_{12}) & 0 & 0 & 0 & x_{12}^{1/3}\,c & 0 & 0 \\
                  0 & 0 & 0 & 0 & 0 & b(x_{12}) & 0 & x_{12}^{-1/3}\,c & 0 \\
                  0 & 0 & 0 & 0 & 0 & 0 & 0 & 0 & a(x_{12}) \\
                \end{array}\nn
              \right)
\end{eqnarray}
with
\begin{equation}
a(x_{12})=(q\,x_{12}-q^{-1}\,x_{12}^{-1}),\qquad
b(x_{12})=(x_{12}-x_{12}^{-1}),\qquad c=(q-q^{-1}),
\end{equation}
and
$$
x_k=e^{h\gamma \theta_k/2}\quad k=1,2, \qquad q=-e^{-i\pi\gamma},
\qquad \gamma=\frac{4\pi}{\beta^2}-1.
$$
Finally, $\rho$  is a scalar function constrained by consistency
relations such as bootstrap constraints, and requirements such as
crossing, which a scattering matrix must satisfy. Its expression can
be found in \cite{Hollowood93}.

\subsection{Transmission matrices: two different approaches}

\p Consider the following static field configurations
\begin{equation}\label{staticconfigurations}
(\phi,\psi)=\frac{2\pi}{\beta} (r,s),
\end{equation}
where $r$, $s$ are any two elements of the root lattice. It is not
difficult to check, looking at \eqref{DefectLag} and
\eqref{DefectTermToda} that, despite having a discontinuity at the
location of the defect, the constant configurations
\eqref{staticconfigurations} all have the same energy and momentum,
namely
$$({\cal E}_0,\, {\cal P}_0)=-\frac{2hm}{\beta^2}(\cosh\eta,\
-\sinh\eta ),$$ and they are the vacuum configurations of the
system. Suppose that a jump-defect is labelled by these vacuum
configurations, in the sense that when the fields $\phi$, $\psi$
have the constant values \eqref{staticconfigurations}, the label
$(r,s)$ is ascribed to the defect. The idea, first presented in
\cite{bczsg05}, is to compare the transmission matrix elements
describing the evolution of the field configurations in the presence
of two different defects: one labelled $(r,s)$ and the other
$(0,0)$. For doing so, the fields $\phi$, $\psi$ are shifted as
follows
$$
\phi\rightarrow \phi-\frac{2\pi r}{\beta}, \qquad
\psi\rightarrow \psi-\frac{2\pi s}{\beta}.
$$
Note that the bulk and the defect potential ${\cal B}$
\eqref{DefectPotentialToda} do not change under this shift, but the
part linear in time derivatives appearing in the defect term
\eqref{DefectTermToda} does. As a consequence, the functional
integrals, which represent the transmission factors related to the
two differently labelled defects, will differ by a constant amount,
namely
\begin{equation}\label{TF1}
T(r,s)=e^{i\tau(r,s)}\, T(0,0),
\end{equation}
where
$$
\tau(r,s)=\frac{\pi}{\beta}\left(-\delta\phi\cdot
(Er+Ds)+(rD+sE)\cdot\delta\psi\right),
$$
and $\delta\phi,\ \delta\psi$ are the changes in the field
configurations from initial to final states. To obtain explicit
expressions for the elements of the soliton transmission matrix for
the $a_2$ model, consider that a soliton passing the defect will
either retains its topological charge or change it to one of the
other weights $l_k$ listed in \eqref{weightsa2}. Therefore, the
effect of a soliton passing the defect must be to change the defect
labels by
\begin{equation}
r\rightarrow r-l_i,\qquad s\rightarrow s-l_j,
\end{equation}
which implies
$$
\delta\phi=-\frac{2\pi l_i}{\beta}, \qquad \delta\psi= -\frac{2\pi
l_j}{\beta} \qquad i,j=1,2,3.
$$
Consequently, expression \eqref{TF1} becomes
\begin{equation}\label{TF2}
T(r,s,l_i,l_j)=e^{i\tau(r,s,l_i,l_j)}\, T(0,0,l_i,l_j)
\end{equation}
where
$$
\tau(r,s,l_i,l_j)= \frac{2\pi^2}{\beta^2}\left(l_i\cdot
(Er+Ds)-(rD+sE)\cdot l_j\right).
$$
In the end, the functional integral approach suggests the following
form for the elements of the transmission matrix (see
\cite{Corrigan07} for details)
\begin{equation}\label{T1}
T{_{i\alpha}^{j\beta}}(\theta)= Q^{\,\alpha\cdot
[E(l_i-l_j)+l_i+l_j]/2} \, T{_{i}^{j}}(\theta)\,
\delta_{\alpha}^{\beta-l_i+l_j},
\end{equation}
where
$$
\alpha=s-r, \qquad Q=-e^{i\pi\gamma}.
$$
Note that the matrix \eqref{T1} is infinite dimensional with roman
and greek labels denoting soliton states and defect charges,
respectively. Naturally, this kind of argument does not provide any
information concerning the rapidity dependent part of the
transmission matrix \eqref{T1}. However, important information
concerning this unknown quantity can be collected making use of a
bootstrap procedure.

\p Consider $D_{\alpha}$ to be the defect operator. Then, it is
formally possible to describe the interaction between a defect and a
soliton or antisoliton as follows ($\theta>0$),
\begin{equation}\label{defectoperator}
A_{i}(\theta)D_{\alpha}=T{^{j\beta}_{i\alpha}}(\theta)D_{\beta}A_{j}(\theta),\qquad
\bar{A}_{i}(\theta)D_{\alpha}=\bar{T}{^{j\beta}_{i\alpha}}(\theta)D_{\beta}\bar{A}_{j}(\theta)\qquad
i=1,2,3,
\end{equation}
where $A_i$, $\bar{A}_{i}$ are operators representing the soliton
and antisoliton states, respectively. Since the antisoliton states
$\bar{A}_{i}$ can be built making use of the soliton states $A_i$,
the two expressions in \eqref{defectoperator} can be combined
together to provide a link amongst the elements of $T$ and
$\bar{T}$. The constraints obtained allow to fix, up to an overall
scalar function of the rapidity, both the matrices $T$, $\bar{T}$
and, surprisingly, to determine the constraints
\eqref{constraintsonD} that the classical quantity $D=1-E$ has to
satisfy. A complete discussion and explicit calculations are
reported in \cite{Corrigan07}.

\p Before revealing the explicit expressions of the two transmission
matrices, a few words must be said on the alternative approach
mentioned in section (\ref{Quantum domain}). It consists in solving
directly the triangular equations, which relate, for instance, the
elements of the soliton transmission matrix $T$ to the scattering
soliton-soliton $S$-matrix elements. Adopting the same conventions
as before for the roman and greek labels, and considering solitons
travelling along the positive $x$-axis ($\theta_1> \theta_2$), the
triangular equations read
\begin{equation}\label{STT}
S{_{kl}^{mn}}(\theta_{12})\,T{_{n\alpha}^{t\beta}}(\theta_1)\,
T{_{m\beta}^{s\gamma}}(\theta_2)=T{_{l\alpha}^{n\beta}}(\theta_2)\,
T{_{k\beta}^{m\gamma}}(\theta_1)\,S{_{mn}^{st}}(\theta_{12}).
\end{equation}
These equations have been discussed first in the context of purely
transmitting defects by Delfino, Mussardo and Simonetti in
\cite{Delf94}. Making use of the $S$-matrix \eqref{SMatrix} and of
the following ansatz for the transmission matrix elements
\begin{equation}\label{Telements}
T{^{n\beta}_{i\alpha}}(\theta)=
t{_{i\alpha}^{n}}(\theta)\;\delta^{\beta-l_i+l_n}_{\alpha}\qquad
i,n=1,2,3,
\end{equation}
it is possible to classify the solutions of \eqref{STT}. This much
has been done in \cite{Corrigan07}, where it was found that one of
the solutions obtained coincides exactly with the soliton
transmission matrix $T$ conjectured by the functional integral
approach. Some of the other solutions may be related to an
alternative setting for the jump-defect with respect to the one
presented in section (\ref{Classical Toda}), while others do not
seem to be relevant for the jump-defect problem. Details are
available in \cite{Corrigan07}. To summarize, the transmission
matrices for solitons and antisolitons related to the jump-defect
presented in section (\ref{Classical Toda}) are, respectively,
\begin{eqnarray}\label{IIAIIBIICdelta}
&&T{^{n\beta}_{i\alpha}}(\theta)=g(\theta) \, \left(
\begin{array}{ccc}
Q^{\alpha\cdot l_1}\,\delta_{\alpha}^{\beta}%\phantom{aaaaaaaa}
  &\hat{x}^2\,\delta_{\alpha}^{\beta-\alpha_1}%\,\phantom{aaaaaaaaaaa}
  &\hat{x}\,Q^{-\alpha\cdot l_2}
  \, \delta_{\alpha}^{\beta+\alpha_0}\phantom{} \\
   \hat{x}\,Q^{-\alpha\cdot l_3}
  \,\delta_{\alpha}^{\beta+\alpha_1}%\phantom{}
  &Q^{\alpha\cdot l_2}\,\delta_{\alpha}^{\beta}%\phantom{aaaaaaaaa}
  &\hat{x}^2\,\delta_{\alpha}^{\beta-\alpha_2}\\%\phantom{aaaaaaaaa} \\
  \hat{x}^2\,\delta_{\alpha}^{\beta-\alpha_0}%\phantom{aaaaaaaaaa}
  &\hat{x}\,Q^{-\alpha\cdot l_1}
  \,\delta_{\alpha}^{\beta+\alpha_2}%\phantom{}
  & Q^{\alpha\cdot l_3}
  \,\delta_{\alpha}^{\beta}\\%\phantom{aaaaaa}
\end{array}
\right),
\end{eqnarray}
and
\begin{eqnarray}
\bar{T}{^{n\beta}_{i\alpha}}(\theta)=\bar{g}(\theta)\label{IIAIIBIICantisolitonsdelta}
\,\left(
\begin{array}{ccc}
 Q^{-\alpha\cdot l_1}\,\delta_{\alpha}^{\beta}\phantom{}
  \,
  &\hat{x}\,\delta_{\alpha}^{\beta+\alpha_1}%\phantom{aaa}
  &0\\%\phantom{aaaaaaaaaaaa} \\
  0%\phantom{aaaaaaaaaaaa}
  &Q^{-\alpha\cdot l_2}\,\delta_{\alpha}^{\beta}\phantom{}
  &\hat{x}\,\delta_{\alpha}^{\beta+\alpha_2}\\
  %\,\phantom{aa} \\
 \hat{x}\,\delta_{\alpha}^{\beta+\alpha_0}
 %\,\phantom{aa}
  &0%\phantom{aaaaaaaaaaaa}
  &Q^{-\alpha\cdot l_3}
  \,\delta_{\alpha}^{\beta}\phantom{} \\
\end{array}
\right),
\end{eqnarray}
where $$ \bar{g}(\theta)=g(\theta-i\pi/3)\;g(\theta+i\pi/3)
\;(1+\hat{x}^3), \qquad \hat{x}=e^{\gamma(\theta-\Delta)}.
$$
Eventually, the constant $\Delta$ will be related to the Lagrangian
defect parameter $\sigma$ introduced in \eqref{DefectPotentialToda},
but, first, a few comments are in order. First of all, note the
striking asymmetry of $T$ and $\bar{T}$. Classically, there is
little difference in behaviour between solitons and antisolitons,
and in section (\ref{Classical Toda}) is was pointed out that in
either case the jump-defect causes a phase shift. Depending on the
size of this shift, the topological charge of a soliton or
antisoliton passing through the defect could be converted to just
one of the adjacent topological charges. Comparing expression
\eqref{IIAIIBIICdelta} and \eqref{IIAIIBIICantisolitonsdelta} with
the argument of the classical delay \eqref{argdelay}, it can be seen
that $\bar{T}$ provides a good match to the classical situation
because of the presence of zeros in expected positions, while $T$
does not possess the expected zeros corresponding to the classical
selection rule. It appears that in the quantum context a soliton
passing through the defect may change into either of the solitons
adjacent to it, though the classically allowed transition remains
the most probable.

\p It should be pointed out that solutions \eqref{IIAIIBIICdelta}
and \eqref{IIAIIBIICantisolitonsdelta} are related by a bootstrap
procedure, in the sense that starting with a $T$ matrix
\eqref{IIAIIBIICdelta} for the solitons, the bootstrap leads to the
$\bar{T}$ matrix \eqref{IIAIIBIICantisolitonsdelta} for the
antisolitons. Similarly, starting with the antisoliton matrix
\eqref{IIAIIBIICantisolitonsdelta}, the bootstrap leads to the
soliton matrix \eqref{IIAIIBIICdelta}. A different setting for the
jump-defect would present a situation in which the asymmetry of $T$
and $\bar{T}$ is maintained but the role of solitons and
antisolitons is interchanged.

\subsection{The overall scalar function: additional constraints}

\p Some additional requirements are needed to be able to fix the
overall functions of the transmission matrices. They are provided by
crossing
\begin{equation}\label{crossing}
\bar{T}{_{n\alpha}^{i\beta}}(\theta)=\tilde{T}{_{i\alpha}^{n\beta}}(i\pi-\theta),
\end{equation}
which allows to relate the transmission matrix for antisolitons
$\bar{T}$ to the transmission matrix $\tilde{T}$, which represents a
process in which the incoming particles meet the defect from the
right. In the jump-defect problem, parity is explicitly violated and
therefore the matrix $\tilde{T}$ is expected to differ from the
matrix $T$. Nevertheless, the two matrices $T$ and $\tilde{T}$ are
expected to be related by
\begin{equation}\label{unitarity}
T{_{a\alpha}^{b\beta}}(\theta)\,\tilde{T}{_{b\beta}^{c\gamma}}(-\theta)
=\delta^{c}_{a}\delta^{\gamma}_{\alpha}.
\end{equation}
This constraint replaces the usual unitarity condition, which does
not hold here due to the fact that the model investigated is not
unitary. Making use of solutions \eqref{IIAIIBIICdelta} and
\eqref{IIAIIBIICantisolitonsdelta} in \eqref{crossing} and
\eqref{unitarity} leads to a relationship between the functions $g$
and $\bar{g}$, from which the following minimal solution for $g$ is
derived
\begin{equation}\label{gfunction}
g(\theta)=\frac{f(\theta)}{(2\pi)^{2/3}\,\hat{x}} %\equiv g'(\Theta),\nn
\end{equation}
\p with
\begin{equation}
f(\theta)=\Gamma[(1+\gamma)/2-z]\,\,\prod_{k=1}^{\infty}
\frac{\Gamma[(1+\gamma)/2+3k\gamma-z]\,
\Gamma[(1-\gamma)/2+(3k-2)\gamma+z]}
{\Gamma[(1-\gamma)/2+3k\gamma+z]\,
\Gamma[(1+\gamma)/2+(3k-1)\gamma-z]},
\end{equation}
where $z=i3\gamma(\theta-\Delta)/2\pi$. Note the presence of a pole
in \eqref{gfunction} at
\begin{equation}\label{poleQuantum}
\theta_P =\Delta -\frac{i\pi}{3} -\frac{i\pi}{3\gamma}.
\end{equation}
Comparing this in the classical limit, $1/\gamma\rightarrow 0$
($\beta \rightarrow 0$), with the pole \eqref{poleClasSet} appearing
in the classical delay allows a determination of the relationship
between the parameter $\Delta$ appearing in the transmission matrix
and the defect parameter $\sigma$ appearing in the Lagrangian
density. This relationship reads
\begin{equation}\label{delta}
\Delta=\eta+\frac{i\pi}{2},\qquad \sigma=e^{-\eta}.
\end{equation}
The identification \eqref{delta} is also supported by the results
found during the calculation of the transmission factors for the
lightest breathers, as explained in \cite{Corrigan07}. Besides, the
computation of the energy of the state associated with the pole
\eqref{poleQuantum} reveals that it corresponds to an unstable bound
state, provided $\frac{1}{2} <\gamma<2$. Consequently, in the
classical limit, this unstable state disappears completely. This
fact agrees nicely with the classical finding that a soliton with
real rapidity cannot be absorbed by the defect. It is worth pointing
out that the latter phenomenon differs from the sine-Gordon case in
which a soliton can be absorbed by the defect and consequently a
quantum unstable bound state is always present, independently of the
range of the coupling constant.

\section{Conclusion}

\p Recent results in the context of the $a_2$ affine Toda field
theory concerning the existence of a special integrable defect -
called a jump-defect - have been presented. For this model, it was
possible to provide a complete and consistent description both in
the classical and quantum domains. In particular, the interaction
between the soliton solutions of the $a_2$ affine Toda model and a
jump-defect was found to be described, in the quantum context, by
infinite dimensional matrices that are solutions of the triangular
equations. Unfortunately, there was no room to discuss here further
interesting issues, such as the connection with B\"acklund
transformations or the scattering of defects in motion.

\p The jump-defect problem can be extended to all the $a_r$ affine
Toda models. On the other hand, the existence of integrable, purely
transmitting defects in the other ATFTs appears to be more difficult
to prove. In principle, infinite dimensional solutions of the
triangular equations can be found for some other Toda models, but it
remains to be seen if these solutions can be regarded as
transmission matrices.

\vskip 1.0cm \noindent{\bf Acknowledgements} \vskip .25cm \noindent
The author thanks the organizers of the workshop RAQIS07 for gaving
her the opportunity to present this talk. She is indebted to her
colleagues Peter Bowcock and Ed Corrigan for fruitful collaborations
and she thanks especially Ed Corrigan for many useful discussions.
The author was supported by a postdoctoral fellowship SPM 06-13
supplied by the \emph{Centre National de la Recherche Scientifique}
(CNRS).


\begin{thebibliography}{99}

\bibitem{bczlandau} P. Bowcock, E. Corrigan and C. Zambon,
\emph{Classically integrable field theories with defects},  in
Proceedings of the 6th International Workshop on Conformal Field
Theory and Integrable Models, Landau Institute, September 2002, Int.
J. Mod. Physics {\bf A19} (Supplement) (2004) 82.

\bibitem{bczsg05} P. Bowcock, E. Corrigan and C. Zambon, \emph{Some aspects of
jump-defects in the quantum sine-Gordon model}, JHEP {\bf 0508}
(2005) 023.

\bibitem{Corrigan06} E. Corrigan and C. Zambon, \emph{Jump-defects in the nonlinear
Schr\"{o}dinger model and other non-relativistic field theories},
Nonlinearity {\bf 19} (2006) 1447

\bibitem{Arinshtein79} A. E. Arinshtein, V. A. Fateev and A. B.
Zamolodchikov, \emph{Quantum S-matrix of the $(1+1)$-dimensional
Toda chain}, Phys. Lett. {\bf B87} (1979) 389; \jump
\newline A. V. Mikhailov, M. A.
Olshanetsky and A. M. Perelomov, \emph{Two-dimensional generalized
Toda lattice}, Comm. Math. Phys. {\bf 79} (1981) 473; \jump
\newline G. Wilson,
\emph{The modified Lax and two-dimensional Toda lattice equations
associated with simple Lie algebras}, Ergod. Th. and Dynam. Sys.
{\bf 1} (1981) 361.

\bibitem{Corrigan94} E. Corrigan, \emph{Recent developments in affine Toda
quantum field theory}, published in \emph{Particle and Fields}
(Banff, AB, 1994) CRM Ser. Math. Phys. Springer, New York, (1999).

\bibitem{bcztoda} P. Bowcock, E. Corrigan and C. Zambon,
\emph{Affine Toda field theories with defects}, JHEP{\bf 01} (2004)
056.

\bibitem{Corrigan07} E. Corrigan and C. Zambon, \emph{On purely transmitting defects
in affine Toda field theory}, JHEP, {\bf 07} (2007) 001.

\bibitem{Hollowood92} T. J. Hollowood, \emph{Solitons in affine Toda field
theories}, Nucl. Phys. {\bf B384} (1992) 523.

\bibitem{McGhee94} W. A. McGhee, \emph{The Topological
Charges of the $a_n^{(1)}$ Affine Toda Solitons}, Int. J. Mod. Phys.
{\bf A9} (1994) 2645.

\bibitem{Hollowood93} T. J. Hollowood, \emph{Quantizing $Sl(N)$ solitons and the
Hecke Algebra}, Int. J. Mod. Phys. {\bf A8} (1993) 947.

\bibitem{Jimbo89} M. Jimbo \emph{Introduction to the Yang-Baxter equation}, Int. J. Mod.
Phys. {\bf A4} (1989) 3759.

\bibitem{Delf94}
G. Delfino, G. Mussardo and P. Simonetti, \emph{Statistical models
with a line of defect}, Phys. Lett. {\bf B328} (1994) 123; \jump
\newline G. Delfino, G. Mussardo and P. Simonetti, \emph{Scattering
theory and correlation functions in statistical models with a line
of defect}, Nucl. Phys. {\bf B432} (1994) 518.


\end{thebibliography}
\end{document}